\documentclass[preprint]{aastex}

\def\citet#1{\citeauthor{#1} (\citeyear{#1})}

\def\citep#1{\citeauthor{#1}, \citeyear{#1}}

\def\referee#1{{#1}}

\begin{document}

\title{Observations and modeling of 
the early acceleration phase of erupting filaments involved in  
coronal mass ejections}

\author{Carolus J. Schrijver$^1$, Christopher Elmore$^1$, 
Bernhard Kliem$^{2,3}$, Tibor T{\"o}r{\"o}k$^{4,5}$, and Alan M. Title$^1$}
\affil{$^1$ Lockheed Martin Advanced Technology Center,
3251 Hanover Street, Palo Alto, CA 94304, U.S.A; \\
$^2$ Astrophysical Institute Potsdam, An der Sternwarte 16, 14482 Potsdam, 
Germany;\\
$^3$ Kiepenheuer Institute for Solar Physics, Sch{\"o}neckstr. 6,
79104 Freiburg, Germany;\\
$^4$ University College London, 
Mullard Space Science Laboratory, Holmbury St. Mary, Dorking, 
Surrey RH5 6NT, United Kingdom;\\
$^5$ LESIA, Observatoire de Paris, CNRS. Univerit{\'e} Paris Diderot,
5 Place Jules Janssen, 92190 Meudon, France}
\email{schryver@lmsal.com; celmore@lmsal.com; 
bkliem@aip.de; tt@mssl.ucl.ac.uk; title@lmsal.com}

\date{\today}

\begin{abstract}
We examine the early phases of two near-limb filament destabilizations
involved in coronal mass ejections on 16 June and 27 July 2005, using
high-resolution, high-cadence observations made with the Transition
Region and Coronal Explorer (TRACE), complemented by coronagraphic
observations by Mauna Loa and the SOlar and Heliospheric Observatory
(SOHO).  The filaments' heights above the solar limb in their
rapid-acceleration phases are best characterized by a height
dependence $h(t)\propto t^m$ with $m$ near, or slightly above, 3 for
both events. Such profiles are incompatible with published results for
breakout, MHD-instability, and catastrophe models.  We show numerical
simulations of the torus instability that approximate this height
evolution in case a substantial initial velocity perturbation is
applied to the developing instability. We argue that the sensitivity
of magnetic instabilities to initial and boundary conditions requires
higher fidelity modeling of all proposed mechanisms if observations of
rise profiles are to be used to differentiate between them.  The
observations show no significant delays between the motions of the
filament and of overlying loops: the filaments seem to move as part of
the overall coronal field until several minutes after the onset of the
rapid-acceleration phase.
\end{abstract}

\keywords{Sun: coronal mass ejections (CMEs) -- Sun: filaments}

\section{Introduction}
Observations of the early rise phase of filaments and their overlying fields
can in principle help constrain the mechanisms involved in
the destabilization of the magnetic configuration through comparison
with numerical simulations (e.g.,
\citep{fan2005}; \citep{torok+kliem2005}; \citep{williams+etal2005};  
and references therein), because the detailed evolution depends
sensitively on the model details.  For example, a power-law rise with
an exponent $m=2.5$ was obtained for a slender flux tube in the
two-dimensional version of the catastrophe model
(\citep{Priest&Forbes02}). An MHD instability triggered by an
infinitesimal perturbation implies an exponential rise, as was
verified, for example, for a three-dimensional flux rope subject to a
helical kink instability (\citep{Torok&al04},
\citep{torok+kliem2005}). The same holds for the torus (expansion)
instability (TI), which starts as a $\sinh(t)$ function
(\citep{Kliem&Torok06}) that is very similar to a
pure exponential early on.  The CME rise in a breakout model
simulation was well described by a parabolic profile
(\citep{Lynch&al04}).

The early rise phase of erupting filaments is best observed near the
solar limb using high-resolution data, both in space and in time. Such
data can be obtained by, for example, Big Bear Solar Observatory
H$\alpha$ observations (e.g., \citep{kahler+etal1988}), the Mauna Loa
K-coronameter (e.g., \citep{gilbert+etal2000}), the Nobeyama
Radioheliograph (e.g., \citep{gopalswamy+etal2003};
\citep{kundu+etal2004}), and the Transition Region 
and Coronal Explorer, TRACE (e.g., \citep{vrsnak2001};
\citep{gallagher+etal2003}; \citep{goff+etal2005};
\citep{sterling+moore2004}; \citep{sterling+moore2005};
\citep{williams+etal2005}). 
In those few cases where observers had the field of view for an
appropriate diagnostic to attempt to establish whether the high loops
or the filaments were accelerated first, the temporal resolution often
was not adequate (see, e.g., \citep{sterling+moore2004}, who use the
standard 12-min.\ cadence of SOHO/EIT).

These studies show that filaments that are about to erupt often --~but
not always~-- exhibit a slow initial rise during which both the
filament and the overlying field expand with velocities in the range
of $1-15$\,km/s. Then follows a rapid-acceleration phase during which
velocities increase to a range of $100$\,km/s up to over
$1000$\,km/s. The rapid-acceleration phase finally transitions into a
phase with a nearly constant velocity or even a deceleration into the
heliosphere.

The height evolution immediately following the onset of the rapid
acceleration phase is often approximated by either an exponential
curve (e.g., \citep{gallagher+etal2003}; \citep{goff+etal2005};
\citep{williams+etal2005} --~who also show systematic deviations from 
that fit up to 2$\sigma$ in position~-- ) or by a
constant-acceleration curve (e.g., \citep{kundu+etal2004}; and
\citep{gilbert+etal2000} --~ who show one case in which
a third-order curve improves the fit to the earliest evolution, and
leave others for future analysis); \citet{kahler+etal1988} fit curves
for the acceleration $a=ct^b$ to the first $10-50$\,Mm for four
erupting filaments, but do not list the best-fit values.
\cite{alexander&al02} find a best fit for the height of the early
phase of a CME observed in X-rays by YOHKOH's SXT of the form
$h_0+v_0t+ct^{3.7\pm0.3}$. For 184 prominence events observed by the
Nobeyama Radioheliograph, \citet{gopalswamy+etal2003} show that higher
in the corona velocity profiles include decelerating, constant
velocity, and accelerating ones for heights from $\sim 50$\,Mm to
700\,Mm above the solar surface.

In many cases, the detailed study of the evolution of the early
phase is hampered by insufficient temporal
coverage or by gaps between the fields of view of two complementing
instruments that can be as large as a few hundred Mm. This results in
substantial uncertainties in the height evolution. \citet{vrsnak2001},
for example, concludes that ``[t]he main acceleration phase \ldots\ is
most often characterized by an exponential-like increase of the
velocity'', but notes that polynomial or power-law functions fit at
comparable confidence levels.

In this study, we examine two events displaying the early
destabilization and acceleration of ring filaments leading to coronal
mass ejections. The high cadence down to 20\,s, and the high spatial
resolution of 1\,arcsec, for the early evolution result in relatively
small uncertainties in the height profiles. This enables a sensitive
test of the height evolution against exponential, parabolic, and
power-law fits. We find that a power-law with exponent near $3$, or
slightly higher, is
statistically preferred in both cases. As no published model matches
that profile, we experiment with a numerical model for the 
torus instability, and find that this model can indeed approximate the
observations provided that a sufficiently large initial velocity
perturbation is applied (without which an exponential-like
profile would be found). This finding reminds us of the sensitivity of
developing instabilities to both initial and boundary conditions, 
and shows that the models, particularly their parametric
dependencies, need to be worked out in greater detail in order to
use observations of the height-time observations
to differentiate successfully between competing models.
 
\section{Observations}\label{sec:Observations}
Primary data for this study were collected by the Transition Region
and Coronal Explorer (TRACE; see \citep{traceinstrument}), and
ancillary data by the Mauna Loa Solar Observatory Mark IV
K-Coronameter (MLSO MK4) and the SOHO/LASCO C2 and C3 instruments
(\citet{brueckner+etal1995}).  The events we studied occurred on 16
June 2005 19:10\,UT to 20:24\,UT (emanating from NOAA Active Region
10775), and on 27 July 2005 from 03:00\,UT to 06:20\,UT (from
AR\,10792).

\subsection{16 June 2005}
This eruption in AR\,10775 was associated with an M4.0 X-ray flare. 
TRACE data are examined
from 19:10:42 through 20:08:37\,UT; MLSO MK4 data were available
from 20:06:59 to 20:23:25\,UT to characterize the later positions of the
filament. SOHO's LASCO did not observe at this time.
A characteristic TRACE image is shown in Fig.~\ref{fig:1a}, with
a sampling of outlines for filament ridge, loops, and position tracks.

Initial data were taken at 19:10:42, followed by a few frames
beginning at 19:25:32\,UT.  There is a gap in the TRACE data from
19:29:34 to 19:47:35\,UT as the spacecraft traversed a zone of
enhanced radiation in its orbit.  Starting at 19:47:35 each available
image was used for tracking, with a characteristic cadence of
approximately 40\,s, changing with exposure time and depending on data
gaps associated with orbital zones of enhanced background radiation.

As no distinct features could be tracked in the filaments or in the
overlying loop structures, we use outlines of the top segments of the
filament and of some outstanding overlying loops as indicated in
Fig.~\ref{fig:1a}. We assign confidence intervals to these positions
by estimating the range of pixels that provides a reasonable
approximation of a feature.

The rising filament loses a traceable form mid-way through the
acceleration.  Once this occurs, short bright 'streaks' of plasma
parcels show up that are blurred by their motion during the exposures.
The positions of the midpoints of these streaks were used to extend
the position data for the filament rise.  The length of the objects
was estimated by correcting for motion blur estimated from their
displacement from one exposure to the next, and then their average
positions were obtained, complemented by an uncertainty estimate.

The MLSO MK4 data do not provide the same clarity of features to track
as do the TRACE data, and their observations are at a lower spatial
and temporal resolution.  Thus, only an estimate of the filament
position was tracked, and was chosen as the point furthest from the
limb on the innermost feature on each of the images.

The displacement of the approximate outlines was tracked by fitting
parabolas to sets of three adjacent points on each outline. For each
exposure, a vector was computed normal to the approximating parabola
from the central point at time $t_i$ to where it intersects a
subsequent parabolic fit for time $t_{i+1}$. That intersection point
is then used as the central position for the next step in the tracking
algorithm, thus moving from beginning to end in the image sequence.
The track of the filament ridge and of two overlying loops thus
measured are identified in Fig.~\ref{fig:1a}.  The streaks observed in
the later phases were tracked as described above; their positions are
also shown in Fig.~\ref{fig:1a}.

The filament evolves through three stages (Fig.~\ref{fig:1b}): 1) an
initial slow rise phase at a near-constant velocity, followed by 2) a
rapid-acceleration phase, and finally 3) a constant-velocity phase
high in the corona beginning at about 1$R_\odot$ above the surface.

TRACE data for phase 1 up to 19:54:58\,UT show the features to exhibit
an approximately 
constant velocity relative to the solar EUV limb. During this phase,
the filament moves 11,500\,km at an average of 4.4\,km/sec. We note
that the contribution of the solar rotation to this is negligible: for
a filament at geometric height $H(t)$ above the photosphere, the
apparent velocity $\dot{h}$ relative to the solar limb induced by the
perspective change as the Sun rotates is approximated by $\dot{h}
\approx \dot{H} + R_\odot \,\alpha\,\dot{\alpha},$ for a small angle
$\alpha$ between limb direction and current longitude. For $\alpha
\sim 5^\circ$, the apparent motion due to rotation only would be no
more than $0.2$\,km/s, much less than the observed velocity.

The beginning of the acceleration phase was determined by a
combination of visual inspection of the raw images, inspection of
displacement charts, and minimization of $\chi^2$ values for the fits.
These three methods agreed in each case to within a tens of seconds.
The position data were fit with three different functional dependences
of time: a parabolic fit $a+bt+ct^2$, a power law allowing for an
initial rise velocity $a+bt+ct^m$, and an exponential $a+c\exp(dt)$.

The rapid acceleration phase begins at 19:54:58, at which time we note
the initial appearance of a brightening feature across the lower end
of the central barb of the filament.  This time is at the beginning of
a data gap from 19:54:58 to 19:57:36.  The rapid acceleration phase
continues at least until the remnants of the filament leave the TRACE
field of view at 20:09:15.

We find that the rise of the left-hand segment of the filament is best
fit by a power law. The power-law fit is superior to the exponential
fit in the range $2.7 \le m \le 3.9$.  
Fits with $\chi_\nu^2\le 1.31$, i.e., up to the 99\%\ confidence level,
are found for $2.9
\lesssim m \lesssim 3.6$, with a best-fit value of $m=3.25$. 
Setting $m=3$, we obtain $a=21.3\pm0.7$~Mm,
$b=1.7\pm2.8$~km\,s$^{-1}$, and $c=0.25\pm0.01$~m\,s$^{-3}$, with
$\chi_\nu^2=1.05$; if $b\equiv 0$ and $m\equiv 3$, then
$\chi_\nu^2=1.07$, only marginally worse than the best fit.  The best
fit yields a constant jerk of $6c=1.47\pm0.03$~m\,s$^{-3}$.  At the
edge of the MLSO field of view, the velocity approaches a terminal
value of $\sim 750$~km\,s$^{-1}$.

The above near-cubic fit characterizes the data better than the
quadratic or exponential fits ($\chi_\nu^2$ of 4.7 and 2.2,
respectively), and agrees better with the MLSO data for position and
velocity needed farther from the limb.

The initial phase of the destabilization behaves as if the loops and
filament are parts of a rapidly-expanding volume with no discernible
delays between the motions: the separations between the filament ridge
and two loops traced above it (lower dashed and upper solid curves in
Fig.~\ref{fig:1a}) appear to be essentially constant until the field
is disrupted in the mass ejection (see Fig.~\ref{fig:1c}): filament
and high loops destabilize and begin moving at the same time, and the
distance between them stays close to constant.  For both the higher
and slightly lower loops discernible in the upper field, their
distance from the filament is almost unchanged until 20:01:41 for the
outer loop and 19:59:39 for the inner, lower loop.  At this time, the
aggregate distance increases, as the loops begin to move laterally to
the primary motion of the expanding filament quickly.  This indicates
overall that the high field is not evolving substantially to allow
the filament through, as might be expected in, e.g., the breakout process.

\subsection{27 July 2005}
TRACE data for the eruption associated with the 2005/07/27 M3.7 flare
(Fig.~\ref{fig:2a}) were analyzed for 03:00:18 to
04:43:38\,UT. LASCO C2/C3 data of the leading edge of the associated CME
were available from 04:56:37 to
06:18:05\,UT to characterize the later phase. 
This eruption also exhibits three
stages (Figs.~\ref{fig:2b} and~\ref{fig:2c}): 
an initial constant velocity stage, a second rapid
acceleration phase, and a final coasting phase at near-constant
velocity.

The initial slow rise lasts until 04:30:13\,UT.  This rise 
is already underway when TRACE data start at 03:00:18\,UT.  
The early data establish that the
filament and the high field form one slowly expanding system.
The early rise velocity is calculated to be 13.4\,km/sec --~considerably
faster than for the event of 16 June~-- with the filament moving by 17,500\,km 
prior to 04:30:13 when the rapid acceleration phase begins.

The rapid acceleration phase lasts from 04:30:13 to at least
04:43:38\,UT when the filament leaves the TRACE field of view.  The position
data shown in Fig.~\ref{fig:2b} 
through 04:38:53 is for the filament's top ridge, 
and from 04:39:20 to 04:43:21 is
for bright streaks similar to those seen in the 16 June 2005
event.  

The data show this filament is also accelerating with a nearly
constant jerk. The function $h=a+bt+ct^m$ fits the data {very} well
for $2.9\lesssim m \lesssim 3.7$, with $\chi_\nu^2=0.63$ for $m\equiv
3,\, b\equiv 0$.  With $m=3$, the fit yields $a=45.0\pm0.6$~Mm,
$b=4.3\pm3.3$~km\,s$^{-1}$, and $c=0.31 \pm 0.01$\,m\,s$^{-3}$, which
corresponds to a constant jerk of $1.9\pm0.06$~m\,s$^{-3}$. The
essentially cubic fit is also the only one of our fits that reaches the
appropriate height and velocity to follow the leading edge
of the ejection as observed with LASCO C2/C3.  The
exponential fit does not fit the acceleration phase as well
($\chi^2_\nu=2.2$) and makes for a much poorer transition to the LASCO
data past 4:55\,UT.  The velocity for the quadratic fit provides an
even poorer fit to the acceleration phase observed by TRACE
($\chi^2_\nu=9.1$), and it appears far too slow to match the high
transition to constant velocity.

Two arcs in the overlying field were tracked for this event (outlined
in Fig.~\ref{fig:2a}).  The left of the upper field loops is tracked through
04:41:21, and the right loop of the upper field through 04:42:34.
The separation from the rising filament, as
shown in Fig.~\ref{fig:2c}, shows that for both features there is
little to no difference in distance between the filament and loops for
a majority of the early rise and acceleration phases.  At 04:33:34\,UT
the filament begins to approach the right of the upper field loops in
the primary direction of the acceleration, but keeps a constant
distance in the direction normal to the acceleration.  The track on
the right loop, however, was not at the very top, so this is partially
an effect of the filament moving up as the loop moved off to the
side. Once the loop was to the side, the filament could pass by, and
the distance perpendicular to the direction of motion of the filament
remains unchanged.  At the same time, the distance in the direction of
travel of the filament is decreasing between the filament and the left
high loop, but the distance normal to the primary acceleration is
increasing quickly. This increase is most likely due to the track on
the left loop being closer to directly above the filament, so it had
farther to move up and to the side. So, as the filament was moving up,
the left side of the high field significantly displaced in the
direction of the filament's acceleration, as well as normal to the
primary direction of acceleration.

The third stage, seen in the LASCO C2/C3 data, indicates a constant
velocity of $\sim 1,250$\,km/sec.  Although the leading edge always
propagates much faster than the filament in a CME core, the difference
to the velocity at the last of the \textsl{TRACE} data is substantial,
and implies that the acceleration continues at least part
of the way out to the first C2 data at 1.41$R_\odot$.

\section{Comparison with models}
\label{sec:Comparison}
The two filament eruptions analyzed here are best fit by a power-law
height evolution with a power-law index $m$ near 3 or perhaps slightly
higher ($\chi_\nu^2$ values reach unity for values of $m$ of 3.3 and
3.6, respectively; note that these values match the value of
$3.7\pm0.3$ found in the study by \citep{alexander&al02}).  The nearly
constant rate of increase for the acceleration by 1.4--1.9~m\,s$^{-3}$
persists for about $10-15$~min in both events. These phases
were shown to be statistically inconsistent with either a constant
acceleration or an exponential growth.

The jerk values, $d^3h/dt^3$, of 1.4 and 1.9\,m/s$^3$ 
for the two filament eruptions studied here are
very similar. Estimated values using $6\Delta h/ (\Delta t)^3$ based
on erupting filaments up to $\sim 200$\,Mm in the studies referenced
in \S~1 and here range from $\sim 0.3$\,m/s$^3$ (for an M6.5 event
described by \citep{hori+etal2005}, and a C4 event 
observed by \citep{maricic+etal2004} and modeled by \citep{torok+kliem2005})
up to $\sim 50$\,m/s$^3$ (in an X2.5 event described by 
\citep{williams+etal2005}). There is no clear correlation between
flare magnitude and jerk value for the small sample of events,
other than that the largest outlying flare shows the largest outlying
value of jerk (we note that there is also no clear dependence of 
eventual CME speed and flare magnitude --~see \citet{zhang+golub2003}~-- 
although the class of fast CMEs has a 3 times higher maximum X-ray brightness
than the class of slow CMEs). It thus remains unknown what determines
the value of  $d^3h/dt^3$, but the similarity of the values for the two
cases studied here may be fortuitous.

Our observations of two erupting filaments do not match the results of
catastrophe, MHD instability, or breakout models published thus
far. The catastrophe model comes closest with a power-law rise with an
index of 2.5, which is near, but significantly below, the range
readily allowed by the observations.  The simplifying assumptions of a
two-dimensional slender flux rope with unrestricted reconnection below
it may, of course, have modified the height evolution for the
model. Here we explore another effect, namely that of different
initial conditions, specifically for the torus instability (TI). The
TI results if the outward pointing hoop force of a current ring
decreases more slowly with increasing 
ring radius than the opposing Lorentz force
due to an external magnetic field (Bateman, 1978): we investigate
whether the instability can describe the rapid-acceleration phase of
the two events and its transition to a nearly constant terminal
velocity.

The geometry of the two events appears compatible with a torus
instability: the eruption on 16 June 2005 exhibits an expanding main
loop that approaches a toroidal shape within the range observed by
\textsl{TRACE}, and the eruption on 27 July 2005 is consistent with
such a shape seen side-on. Neither shows indications of helical
kinking.  The $\sinh(t)$ profile obtained analytically for the TI by
\cite{Kliem&Torok06} relied on the simplifying assumption
that the external poloidal field varies with the major torus radius
$R$ as $B_\mathrm{ex}\propto R^{-n}$ with a constant decay index $n$,
and it is exact only as long as the displacement from the equilibrium
position remains (infinitesimally) small.

Allowing for a height dependence of the decay index $n$ 
likely will
cause the height evolution in the model to differ even more from the
observations: because $n(h)$ is in reality an increasing function on
the Sun \cite[see, e.g., Fig.~2 in][]{vanTend&Kuperus78}, the
acceleration profile would likely increase more steeply than the 
initially nearly exponential $\sinh(t)$ function. 

We have performed numerical MHD simulations of the TI to study the
evolution for finite displacements. For some parameter settings, the
exponential expansion was found to hold up to several initial radii of
the current ring, while for others a power-law-like expansion with
exponents scattering around $m\sim3$ could be found. The latter turns
out to be related to the influence of the initial velocity on the rise
profile in these MHD simulations. We focus on this aspect below.

Our simulations are largely similar to those of kinking flux ropes in
\cite{torok+kliem2005}, and we refer to that study for model details. 
The flux rope equilibrium by \cite{Titov&Demoulin99} (TD99) is used as
initial condition. The line current in that model, which introduces a
stabilizing external toroidal field is here set to zero, and the
stabilizing influence of line tying is kept small by choosing a torus
center only one tenth of the initial apex height below the bottom plane. In
order to preclude the helical kink instability, a sub-critical twist of
$\Phi\approx2.5\pi$ is chosen, which requires the flux rope to be
relatively thick (the minor radius is
0.6 times the initial apex height, yielding an aspect
ratio of only 1.83). The approximation of a slender flux tube used in
TD99 becomes relatively inaccurate for these settings, so that the
simulations start with a short phase of relaxation toward a numerical
equilibrium, lasting about a dozen Alfv\'en times ($\tau_A$).

The TI is triggered by the motions set up in the relaxation phase,
which may reach one tenth of the Alfv\'en speed ($V_A$, measured at
the flux rope apex in the initial configuration), 
depending on parameters. In a first set of
simulations, we set the decay index of the external poloidal field at
the initial apex height to be $n=1.20$, close to its critical value
analytically derived to be $1.23$ for the parameters given \cite[see
Eq.~(5) in][]{Kliem&Torok06}.  The TI then develops very gradually, in
a period of $\sim100~\tau_A$, while the perturbations caused by the
initial relaxation decay in $\approx15\mbox{--}20~\tau_A$. This
simulation yields a clearly exponential rise profile (solid
lines in Fig.~\ref{fig:TIseries}).

In four subsequent runs in this set, an upward, linearly rising
perturbation velocity is imposed at the flux rope apex of the same
initial configuration at the start of the runs with an increasing
duration (from $6~\tau_A$ up to $10~\tau_A$).
Figure~\ref{fig:TIseries} shows the resulting transition from exponential to
power-law-like rise profiles for these TI simulations.

The fourth run (dashed  lines) approaches a constant-jerk rise
profile best.  This best-fit run has an initial velocity of $0.03~V_A$
at the onset of the TI-driven rise of the acceleration at
$t\approx15~\tau_A$ and $h=1.74$. It approximates constant jerk up to
$t\sim30~\tau_A$ (i.e., nearly until the peak acceleration is
reached) and $h\sim2.8$.

Figure~\ref{fig:constant_jerk_TI} shows another case of close approach
to a cubic rise profile by a flux rope in a strongly torus-unstable
TD99 equilibrium \referee{ (The included scaling of the simulation
data to the rise profile of the 2005/06/16 eruption is discussed
below.)}. {Here the decay index of the external field at the initial apex
height is strongly supercritical, $n=2.85$, close to the asymptotic
value for a dipole field ($n\to3$) in the TD99 equilibrium 
(Figure~\ref{fig:simseq} shows a rendering of this simulation). On the
other hand, with the depth of the torus center chosen to be 3/8 of the
initial apex height, the line tying has a stronger stabilizing
effect. Except for a somewhat larger aspect ratio of 2.3, the other
parameters are identical to those of the runs shown in
Fig.~\ref{fig:TIseries}.} The initial velocity at the onset of the
TI-driven rise of the acceleration is, again, approximately
$0.03~V_A$. This velocity results from the initial, more vigorous
relaxation towards a numerical equilibrium and from the early onset of
magnetic reconnection in the vertical {current sheet, which is formed
below the flux rope similar to the simulation shown in
\cite*{Kliem&al04}. By the end of the relaxation
($t\approx10~\tau_A$), both upward and downward reconnection outflow
jets from the current sheet are formed and the upward jet reaches
$0.03~V_A$, reducing the marked decrease of the upward perturbation
velocities observed in the first run in
Fig.~\ref{fig:TIseries}. During the whole phase of nearly
constant-jerk rise, the TI-driven rise of the flux rope apex and the
upward reconnection outflow jet grow synchronously, reaching similar
velocities.

Such close coupling between the ideal instability and reconnection can
obviously support a power-law rise of the unstable flux rope, but it
is neither a necessary nor a sufficient condition for its occurrence,
as the comparison with Run~4 in Fig.~\ref{fig:TIseries} and with the
CME simulation in \cite{torok+kliem2005} shows. Run~4 exhibits a
nearly power-law rise, but reconnection outflow jets from the vertical
current sheet develop here only after the acceleration of the flux
rope has passed its peak ($t>40~\tau_A$). The CME simulation in
\cite{torok+kliem2005} showed a similar coupling between the ideal MHD
instability (the helical kink in this case) and reconnection as the
run shown in Fig.~\ref{fig:constant_jerk_TI}, but with an initial
velocity of $\approx0.01~V_A$ the rise was clearly exponential.

While all data in Fig.~\ref{fig:TIseries} and the solid line in
Fig.~\ref{fig:constant_jerk_TI} monitor the apex of the magnetic axis
of the flux rope, the dashed lines in Fig.~\ref{fig:constant_jerk_TI}
show the rise of a fluid element near the bottom of the flux rope,
\referee{which is a likely
location for the formation of filaments. Lying initially at $0.65\,h_0$,
it belongs to an outer flux surface of the rope.} 
Although the flux rope in the simulation
expands during the rise, both the axis and the bottom part show an
approximately constant jerk, and no significant timing differences
between the acceleration profiles. 

\subsection{Scaling simulation to observation}
\label{ssec:Scaling}

\referee{
Figure~\ref{fig:constant_jerk_TI} presents a scaling of the simulation
data to the rise profile of the 2005/06/16 eruption, determined in
three steps. First, the time of the velocity minimum near $10\,\tau_A$
in the simulation is associated with the onset time, $t_0$, of the
rapid-acceleration phase, 19:54:58~UT, as obtained in Sect.~2.
Second, the time $t_1$ of maximum simulated velocity is associated
with the time halfway between the final MLSO data points, which yields a
substantially better match between the acceleration profiles than
assuming that the acceleration ceased at or after the final MLSO data. 
These two
choices yield $\tau_A=32.5$~sec. Third, the simulated and observed
heights are matched at $t_1$, resulting in a length unit for the
simulation of $h_0=44.4$~Mm, an Alfv\'en speed
$V_A=h_0/\tau_A=1370$~km/s, and a normalization value for the
acceleration of $a_0=V_A/\tau_A$. Figure~\ref{fig:constant_jerk_TI}
shows the observed heights on a linear scale, with derived velocity
and acceleration data (based on central differences, with a 7-point
boxcar averaging to smooth the heights and velocities, and a 5-point
boxcar averaging for the accelerations).}

\referee{
Both the rise of the magnetic axis of the flux rope (solid line in
Figure~\ref{fig:constant_jerk_TI}) and the rise of a fluid element
originally below the magnetic axis (dashed line) are scaled to the
data. The lower fluid element yields the best match, and is shown in
Fig.~\ref{fig:constant_jerk_TI}.}

\referee{
We note that a correction of the observed heights for perspective
foreshortening may improve the fit of model to observations.  The
\textsl{TRACE} images suggest that the direction of ascent may have
been inclined from the vertical direction by $\sim45^\circ$ at the
onset of the accelerated rise, and it is plausible to assume that it
had become vertical by the time of the final MLSO data point. Such a
correction brings all height data points even closer to the dashed
line in Fig.~\ref{fig:constant_jerk_TI}. However, since such a
correction introduces a degree of uncertainty while the effects are
relatively minor, we do not attempt to apply such a correction. }

\referee{
Not only is the overall match between the observations
and the scaled simulation quite satisfactory, 
the scaling also yields plausible values for the Alfv\'en velocity and the
footpoint spacing of the model flux rope, $D_\mathrm{foot}=98$~Mm. The
latter agrees well with the observed value, which Fig.~1 suggests
to be $\approx94$~Mm (from $x\approx230$ to $x\approx490$), or slightly
larger owing to foreshortening. The TD99 equilibrium
by construction tends to yield a systematically large initial apex
height, so that the footpoint distance provides a far better check of the
length scale when the model is confronted with observations.}

\referee{
The scaling also shows that the filament velocity at the onset of the
rapid rise (first data points after 19:55~UT) nearly reaches the value
of $\sim0.03\,V_A$ required in the simulations of
Figs.~\ref{fig:TIseries} and \ref{fig:constant_jerk_TI} for the
transition from an exponential to a nearly cubic height-time profile. The
observed velocities at $t_0$ 
even exceeded 
the initial velocity of $0.017\,V_A$ of the run shown dotted in
Fig.~\ref{fig:TIseries}, which developed an intermediate rise profile
quite close to the observed profile. We infer from this
that the initial velocity is a parameter which helps
control the detailed properties of the rise profile.}

\referee{
The observations of the eruption on 2005/07/27 do not constrain the
scaling of the simulation as well as the 2005/06/16 data. The LASCO data
at large distances refer to the leading edge of the CME, i.e., to a
different part of the ejection than the \textsl{TRACE} data, and the two
sets do not join to form as nearly a continuous $h(t)$ profile as the
2005/06/16 data. Only the \textsl{TRACE} data can be used for the
scaling, leaving more ambiguity in the scaling for this event. The best
match between the simulation and the data is obtained when the final
\textsl{TRACE} height measurement is assumed to lie slightly past the
time of peak acceleration, by $2\mbox{--}5\,\tau_A$. Equating the
simulated and observed heights at this time gives a match of comparable
quality to the one in Fig.~\ref{fig:constant_jerk_TI} for both the
magnetic axis and the lower fluid element. We present the former in
Fig.~\ref{fig:scaling20050727}, which yields the scaled parameters
$\tau_A=26$~sec, $V_A=940$~km/s, and $D_\mathrm{foot}=55$~Mm. Scaling the
rise of the lower fluid element to the observations yields
$\tau_A=29$~sec, $V_A=1500$~km/s, and $D_\mathrm{foot}=99$~Mm instead. As
with the 2005/06/16 data, the observed velocity closely approaches the
scaled simulation velocity shortly after the estimated onset time of
the fast rise (within $\sim5\,\tau_A$).}

\referee{
The scalings support the hypothesis that the torus instability of a
flux rope has been a possible driver of both eruptive filaments in
their rapid-acceleration phase. We note that the only parameter that
was adjusted particularly to fit the observations is the decay index
for the overlying field ($n=2.85$), since both eruptions evolved into
a moderately fast CME and the TI requires $n\gtrsim 2$ to produce a
fast ejection (\citep{Torok&Kliem2007}).}

\subsection{Dynamics of overlying loops}
\label{ssec:OverlyingLoops}

\referee{
Figure~\ref{fig:TI_4frame_rendering} shows that field lines that
initially pass over the legs of the flux rope, lean strongly sideways
during the rope's rapid acceleration phase, similar to the motion of
the observed overlying loops. Their lateral motions in Figs.~3 and\,6
commence with little or no delay to the beginning rapid acceleration
of the filament (except for a much weaker lateral motion of the left
overlying loop in the slow rise phase of the 2005/07/27 event), and
they combine with the vertical motions such that the total distance
between loop apex and filament apex varies only little in the first
$\approx5$ minutes of the rapid-acceleration phase (corresponding to
$\sim10\,\tau_A$), but increases rapidly thereafter.}

\referee{
We emphasize that the observations of the two events do not permit us
to determine the delay between the start of the displacement of the
overlying loops relative to the filament's rapid acceleration to
better than an Alfv\'en travel time: the Alfv{\'e}n velocities of
order 1,000\,km/s and the instrument cadence mean that signals can
propagate between the overlying loops and the filament within 1 to 2
imaging intervals. Consequently, we can only conclude that the data
are compatible with a delay of at most one Alfv\'en travel time.}

\referee{
Figure~\ref{fig:OverlyingLoops} plots the distances for a set of loops
in a format similar to Figs.~3 and~6. These loops were selected such
that their apex points have equally-spaced initial distances on a
straight line from the origin, inclined by $25^\circ$ from the
vertical. The second lowest of these loops is marked by an asterisk in
Fig.~\ref{fig:TI_4frame_rendering}. We find that the model's
horizontal and vertical distances combine to a slowly varying total
distance for about $10\,\tau_A$ after TI onset (at
$t\approx10\,\tau_A$), followed by a rapid increase of the total
distance, as in the observations.  This behavior occurs in an angular
range between the vertical and the initial origin-apex line of,
roughly, 20--35$^\circ$.  For larger inclinations of the overlying
loop the initial ratio of vertical and horizontal distance is smaller
than observed, and for smaller inclinations the horizontal motion
commences too late.}

\referee{
Figure~11 also reveals two types of perturbations in this simulation. The
first is an initial phase of relaxation from the analytical TD99 field to
a nearby, numerically nearly potential-field state, which occurs in the
whole surrounding field of the flux rope and is of nearly uniform
duration of $2\mbox{--}3\,\tau_A$. The second is a wave-like
perturbation, launched by the (more vigorous) initial relaxation of the
current-carrying flux rope, of duration $\sim10\,\tau_A$, and propagating
outward trough the whole box at about the Alfv\'en speed. The motion of
the overlying loops is seen to commence with the passage of the second
perturbation, i.e., with a delay of only one Alfv\'en travel time, and to
continue smoothly after its passage (similar to the behavior of the flux
rope, whose instability develops out of the initial relaxation). A delay
this short is consistent with the observations.}

\referee{
The feature of an initially only slowly varying total distance occurs in
a substantial height range, so that one cannot conclude that the observed
overlying loops give a good indication of the edge of the flux rope in
the two events considered. However, with increasing initial height of the
loops, the phase of rapid increase of the distance to the rope occurs
progressively delayed. The scalings place the observed transition between
the two phases at $t\sim20\,\tau_A$, in agreement with the lowest two or
three loops included in Fig.~\ref{fig:OverlyingLoops}, indicating that
the overlying loops were located in the range between the surface of the
flux rope and about three minor radii from its axis.}

\section{Conclusions}
\label{sec:conclusions}
We study two well-observed filament eruptions, and find that their
rapid acceleration phases are well fit by a cubic height-time curve
that implies a nearly constant jerk for $10-15$\,minutes, followed by
a transition to a terminal velocity of $\sim 750$\,km/s and $\sim
1250$\,km/s, respectively. Simulations of a torus instability (TI) can
reproduce such a behavior, provided that a substantial initial
velocity perturbation is introduced. Without that perturbation, an
exponential rise profile would be found.

We note that the initial slow rise and the onset of the subsequent rapid
acceleration phase are shared between the filament and overlying loop
structures: neither \referee{leads the other to within the temporal
resolution. For characteristic
Alfv{\'e}n speeds over active regions of $\sim 1,000$\,km/s, the
propagation of a perturbation over the separation of $\sim 75,000$\,km
would require only $\sim 1.2$\,min., which corresponds to only one or
two exposures. Thus the observations allow for Alfv{\'e}nic propagation
of a signal between filament and overlying loops, but suggest no longer-term
differential evolution.}

We observe no significant changes in the separation of erupting
filament and overlying loops within that interval (Figs.~\ref{fig:1c}
and~\ref{fig:2c}). After that, the distance increases in the
2005/06/16 eruption, suggesting the overlying field moves to the side
for some time faster than the filament rises.  For the 2007/07/27
eruption, the distance stays the same for one loop and decreases for
another for up to 10~min after the start of the rapid acceleration
phase, which reflects the significant sideways motion component of the
rising filament. The observed configuration of the filament and high
loops may be part of a larger overall destabilizing field
configuration. Our numerical modeling has assumed that, in the rapid
acceleration phase, the overlying field starts to move rapidly only
as a consequence of the flux rope's destabilization. This is
consistent with the data. However, we cannot exclude that the filament
and the overlying field were destabilized simultaneously by a process
different from the one considered here. More study is needed
to establish whether the common evolution of the filament and high
loops has a significant diagnostic value as to the cause of the
instability.

Comparison with other model studies in the literature leads us to
conclude that the catastrophe model and the TI model are both
marginally consistent with the observations of the two erupting
filaments. The catastrophe model predicts a power-law
exponent near the lower edge of the range of acceptable fits, but we
have to allow for the possibility that changing that model's details
may change the acceleration profile.  
\referee{In order to yield the observed nearly cubic power-law rise (with
$m$ slightly exceeding 3), our TI model requires an initial
perturbation velocity that is in agreement with the observed rise
velocity at the onset of the rapid-acceleration phase.  If a nearly
exact cubic rise were to be matched, however, initial velocities
moderately exceeding the observed ones, by a factor $\approx1.5$, were
required. In any case, our modeling is consistent with the observed
velocities after the first few minutes of the eruption.}

Having established that the model for the TI instability is very
sensitive to the initial conditions, we should of course also
acknowledge that it depends sensitively on the model details
itself. These include the details of the external field and of the
rates and locations of the reconnection that occurs behind the
erupting filament.  That such reconnection occurs in reality is
suggested for both events by the occurrence of brightenings mainly at
the bottom side of the filaments at the onset of the
rapid-acceleration phase. These brightenings develop later into the
streaks used for position determination in
Sect.~\ref{sec:Observations}.  The onset of reconnection even before
the rapid-acceleration phase of the filament eruption on 27 July 2005
is strongly suggested by precursor soft and hard X-ray emission during
about 04:00--04:30~UT, whose analysis revealed heating to 15~MK and
the acceleration of non-thermal electrons to energies $>10$~keV
(\citep{Chifor&al06}).

The observed rise velocity early in the filament eruption may 
be an underestimate of the true expansion
velocity of the hoop formed by the flux rope: the filament channel in
the pre-eruption phase of AR\,10775 is strongly curved, and one of the
two possible channels in AR\,10792 is too (ambiguity exists here
because the eruptions occurred very near the limb, so that the
configurations of the filament channels can only be observed some days
before and after the events, respectively). If the initial expansion
of the flux rope would have a strong component in the general
direction of the inclined plane of the curved filament channel rather
than be purely normal to the solar surface, projection effects could
cause us to underestimate the expansion velocity in particular early in the
evolution.  In addition to that, we must realize that the TI model
assumes a flux rope that stands normal to the solar surface and that
erupts radially. Future more detailed modeling will have to show how
deviations from that affect the evolution of the eruption.

The fact that the torus-instability model yields qualitatively
different rise profiles (exponential vs.\ power law) in different
parts of parameter space, cautions against expectations that precise
measurements of the rise profile of filament eruptions by themselves
permit a determination of the driving process: the non-linearities in
the eruption models clearly require high-fidelity modeling if such
observations are to be used to differentiate successfully between
competing models. Our initial modeling discussed here suggests
that the torus instability is a viable candidate mechanism for at 
least some filament eruptions in coronal mass ejections.
\referee{Given the dependence of nonlinear models on the details 
of boundary and initial conditions, it will be necessary to investigate
how other models for erupting filaments compare to the data, as well as
how the fidelity of our modeling of the torus instability can be improved
before we can reach definitive conclusions about the mechanism(s) responsible
for filament eruptions in general.}

\acknowledgements 
We thank Joan Burkepile for providing us with MLSO MK4 observations,
and Terry Forbes for helpful discussions.  We are grateful to the
referee for constructive and helpful comments that led us to pursue
the model-observation parallels in this study in more detail.  This
work was supported by NASA under the TRACE contract NAS5-38099 with
NASA Goddard Space Flight Center, by NSF grant ATM 0518218 to the
University of New Hampshire, by the European Commission through the
SOLAIRE Network (MTRN-CT-2006-035484), and by the Deutsche
Forschungsgemeinschaft.


\vfill\eject
\section*{Figure captions:}

Figure~\ref{fig:1a}: TRACE 171\,\AA\ image taken at 2005/06/16 19:25:32\,UT. 
Sample outlines of the top edge of the rising filament over time
and of two overlying loop
structures are shown for the time interval from 19:25\,UT to 20:04\,UT.
The positions for which the heights are shown in Fig.~\ref{fig:1b} are marked. 

Figure~\ref{fig:1b}: Distances from the solar EUV limb for the
rising and erupting filament on 2005/06/16 
(see Fig.~\ref{fig:1a} for the tracked
positions). The bottom panel shows the central phase with rapid
filament acceleration in detail. Both panels show several 
fits to the data (see legend). The top panel also shows positions
derived from the MLSO coronagraphic data for the later phase when the eruption
turns into a proper mass ejection. 

Figure~\ref{fig:1c}: Distances between the tracked ridge of the 
filament shown in Fig.~\ref{fig:1a} and the lower (top panel) 
and upper (bottom panel) overlying loops. The total distance is
shown by the solid line; distances in the figure's $x$ and $y$ directions
are shown separately 
by dashed and dashed-dotted lines, respectively. 

Figure~\ref{fig:2a}:  TRACE 171\,\AA\ image taken at 2005/07/27 03:00:08\,UT.
This figure, similar to Fig.~\ref{fig:1a}, identifies segments of
overlying loops on the left and right side of the rising filament. 

Figure~\ref{fig:2b}: As Fig.~\ref{fig:1b} for the event observed
on 2005/07/27. Note that the exponential and quadratic fits
are shown offset by +2\,min. in the top panel to reduce overlap, 
but shown properly placed in time in the lower panel.

Figure~\ref{fig:2c}: Distances between the tracked ridge of the 
filament shown in Fig.~\ref{fig:2a} and the left (top panel) 
and right (bottom panel) overlying loops.

Figure~\ref{fig:TIseries}: 
Transition from exponential to approximately power-law rise profile
with increasing initial velocity for a torus-unstable flux rope
equilibrium with an external field decay index of $n\ge 1.2$ (see text
for other parameter values).  Apex height $h(t)$, velocity $u(t)$,
acceleration $a(t)$, and jerk $j(t)=da/dt$ are normalized using the
initial apex height $h_0$, the Alfv\'en speed $V_A$, and the
corresponding derived quantities. Time is normalized by
$\tau_A=h_0/V_A$.  Solid lines show the unperturbed run, i.e., the
development of the instability from rest.  For the further runs of the
series a velocity perturbation at the apex is linearly ramped up until
6, 8, 9.25, and 10~$\tau_A$ (dashed-dotted, dotted, dashed,
dashed-triple dotted, respectively).

Figure~\ref{fig:constant_jerk_TI}: 
Nearly constant-jerk rise profile for an unperturbed torus-unstable
flux rope equilibrium with steeper field decrease above the flux rope
than in Fig.~\ref{fig:TIseries}; the field decay index in this case is
$n\ge 2.85$, i.e., near the value for the far field in the dipolar case
(see text for other parameter differences for aspect ratio and initial
torus depth).  Solid lines show the rise profile of the apex point of
the magnetic axis as in Fig.~\ref{fig:TIseries}, dashed lines show the
rise profile of a fluid element below the apex, initially at
$h=0.65\,h_0$. \referee{ The simulation data for this lower fluid element are
scaled to the rise profile of the 2005/06/16 filament eruption, and
the resulting Alfv\'en time, Alfv\'en speed, and footpoint distance
are given.}

Figure~\ref{fig:simseq}: Side view of a torus instability simulation
(see Fig.~\ref{fig:constant_jerk_TI}). The field lines of the torus
are shown lying in a flux surface at half the minor torus
radius. Sample field lines for the overlying field are also shown. The
starting points in the bottom plane for the traced field lines are the
same for all panels. The times (expressed in Alfv\'en crossing times,
as in Figs.~\ref{fig:TIseries}--\ref{fig:OverlyingLoops}) are
$=0,\,20,\,30$, and 40, respectively. \referee{The motion of the loop apex
marked by an asterisk is shown in Fig.~\ref{fig:OverlyingLoops}.}

Figure~\ref{fig:scaling20050727}:
\referee{Scaling of the simulation data from Fig.~\ref{fig:constant_jerk_TI} to
the rise profile of the 2005/07/27 filament eruption; here the rise of
the magnetic axis' apex point (solid line) is scaled.}

Figure~\ref{fig:OverlyingLoops}:
\referee{Distances of the apex point of representative loops, initially
overlying the flux rope at an angle of $25^\circ$ from the vertical,
to the lower fluid element of the simulation shown in
Figs.~\ref{fig:constant_jerk_TI} and \ref{fig:scaling20050727} (dashed
line in these figures). The format is similar to Figs.~3 and~6. For
clarity, horizontal and vertical distances are included only for the
lowest and highest of the selected loops.  The second
lowest of these loops is marked by an asterisk in 
Fig.~\ref{fig:TI_4frame_rendering}.}

\vfill\eject

\begin{figure}[ht!]
\epsscale{.65}
\plotone{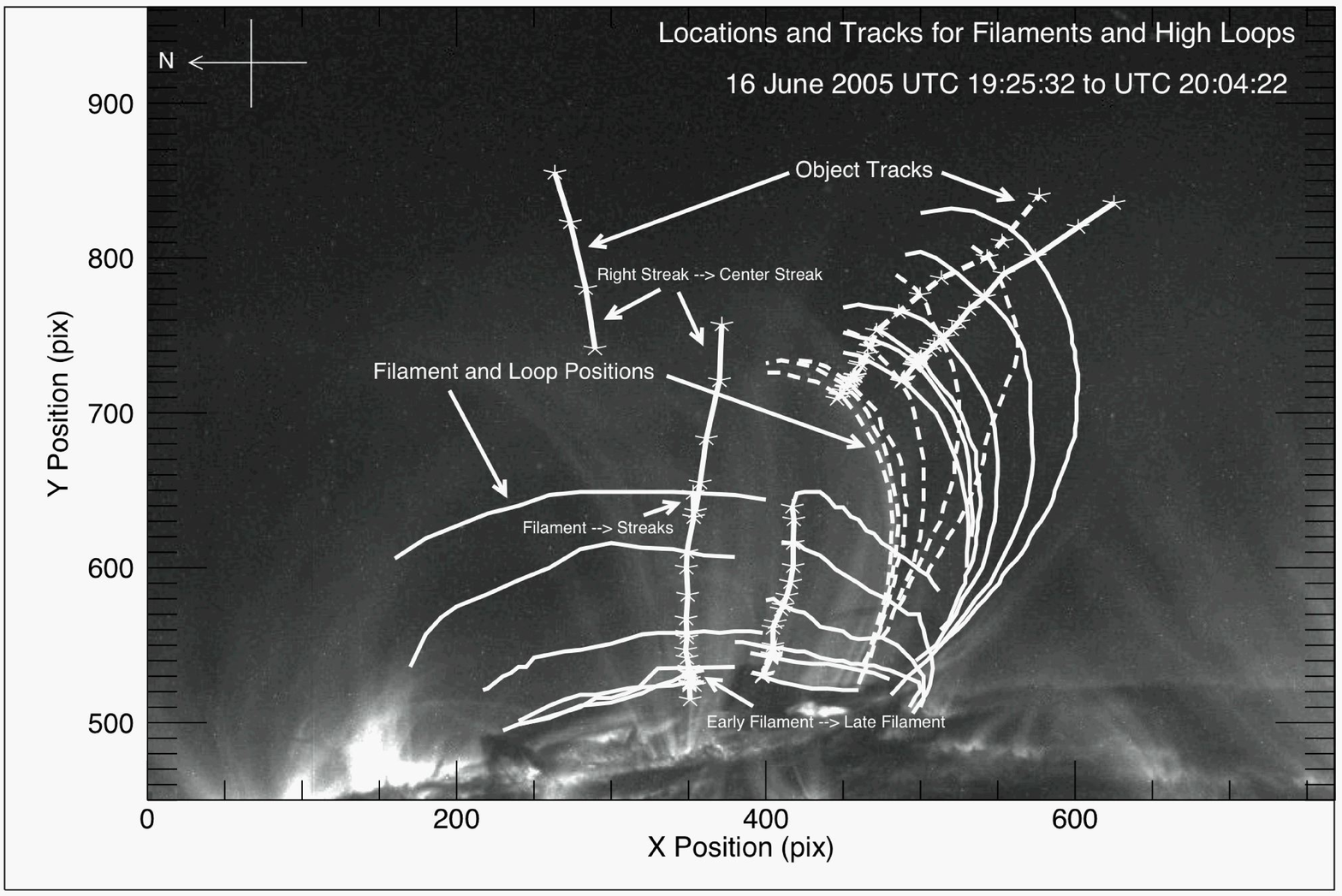}
\caption{\em }\label{fig:1a}
\end{figure}

\begin{figure}[ht!]
\epsscale{.65}
\plotone{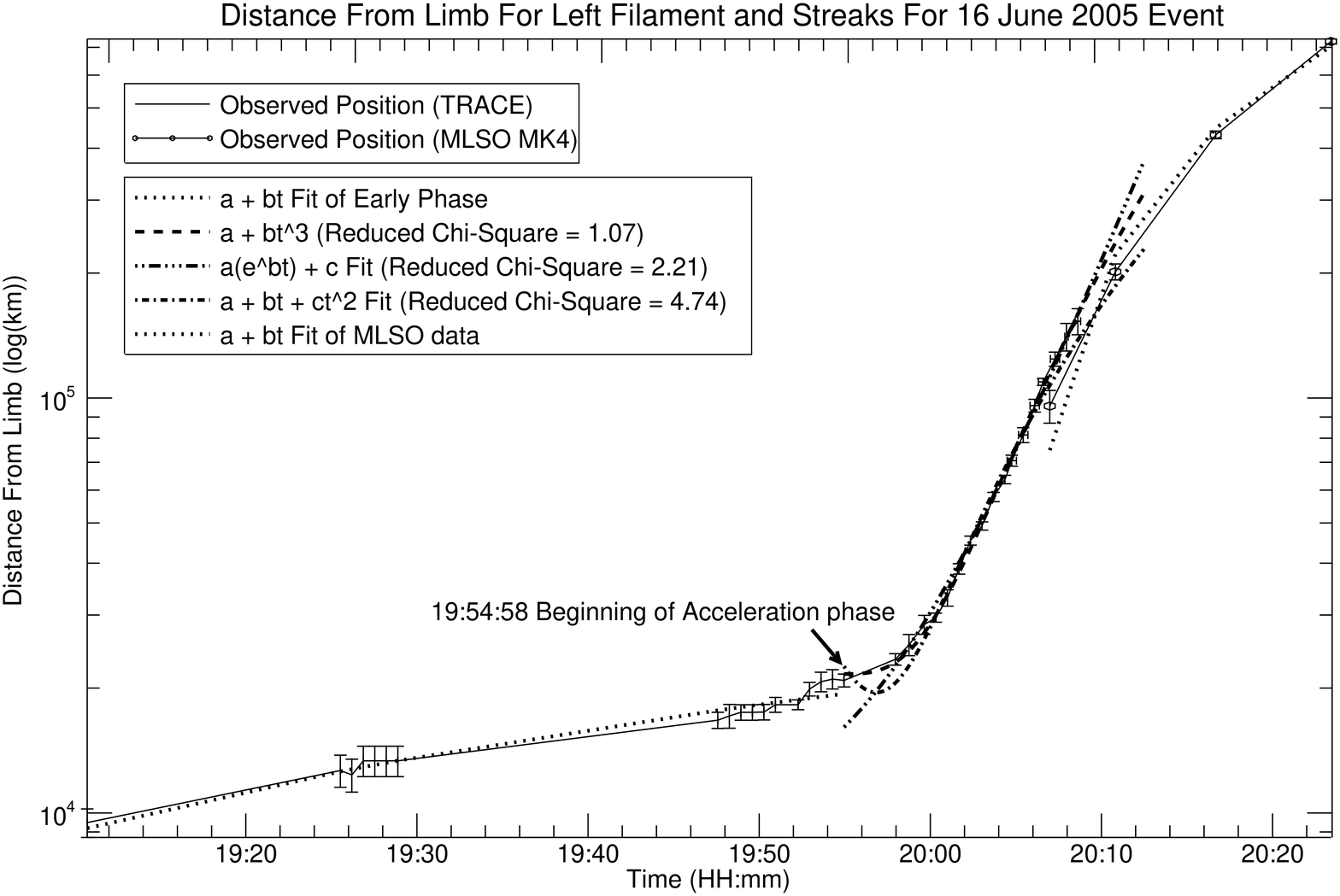}
\plotone{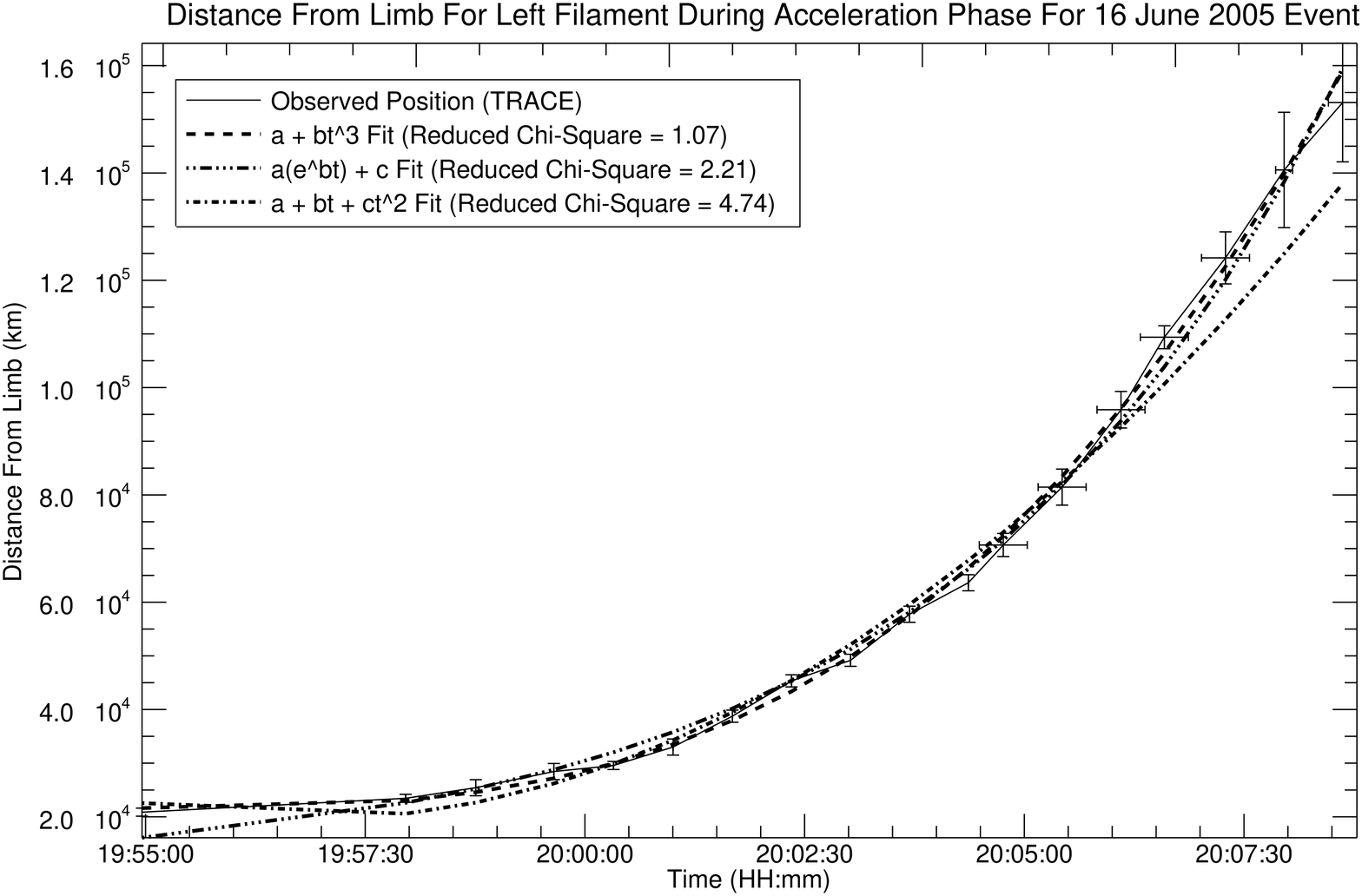}
\caption{\em }\label{fig:1b}
\end{figure}

\begin{figure}[ht!]
\epsscale{.65}
\plotone{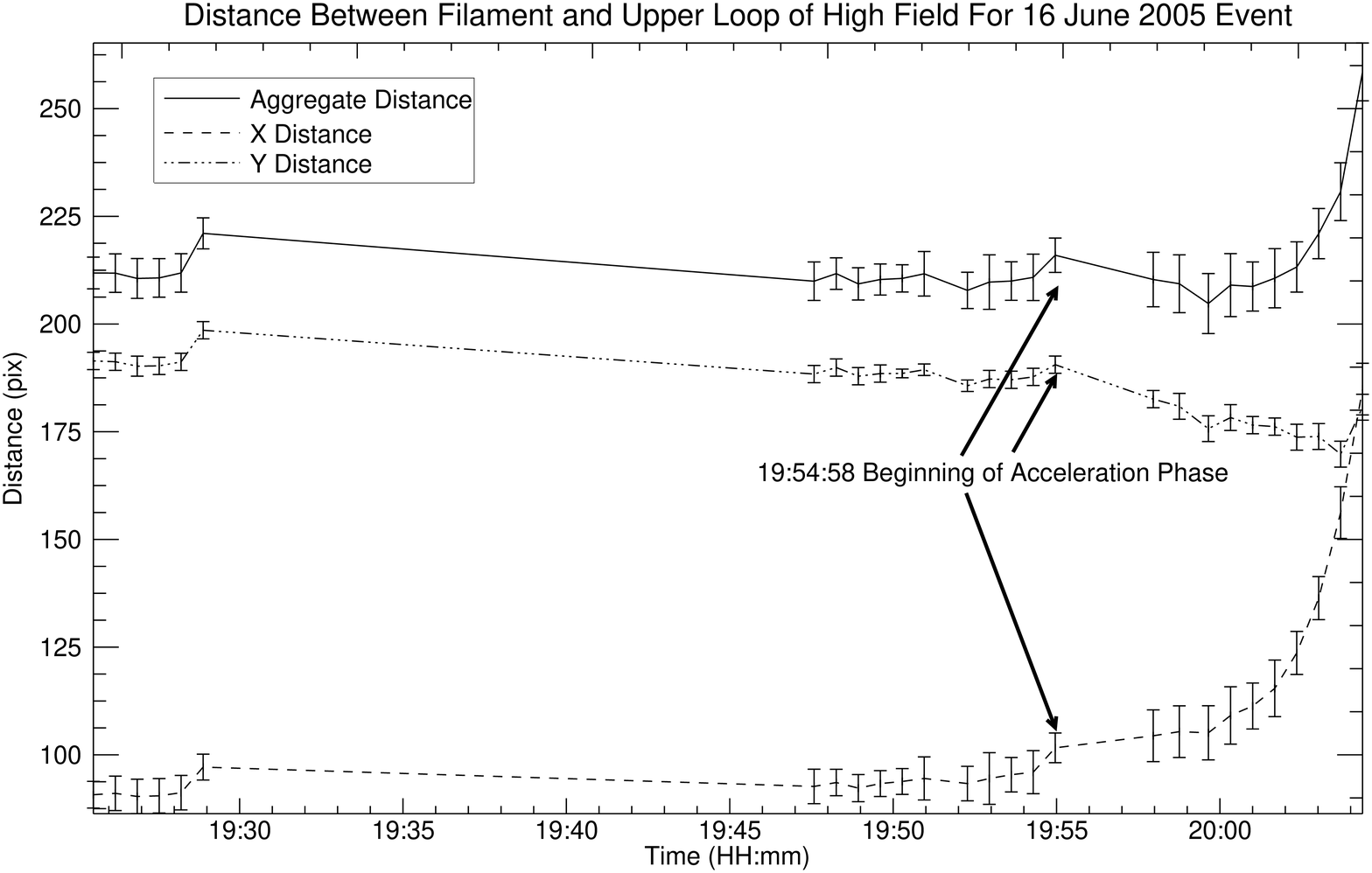}
\plotone{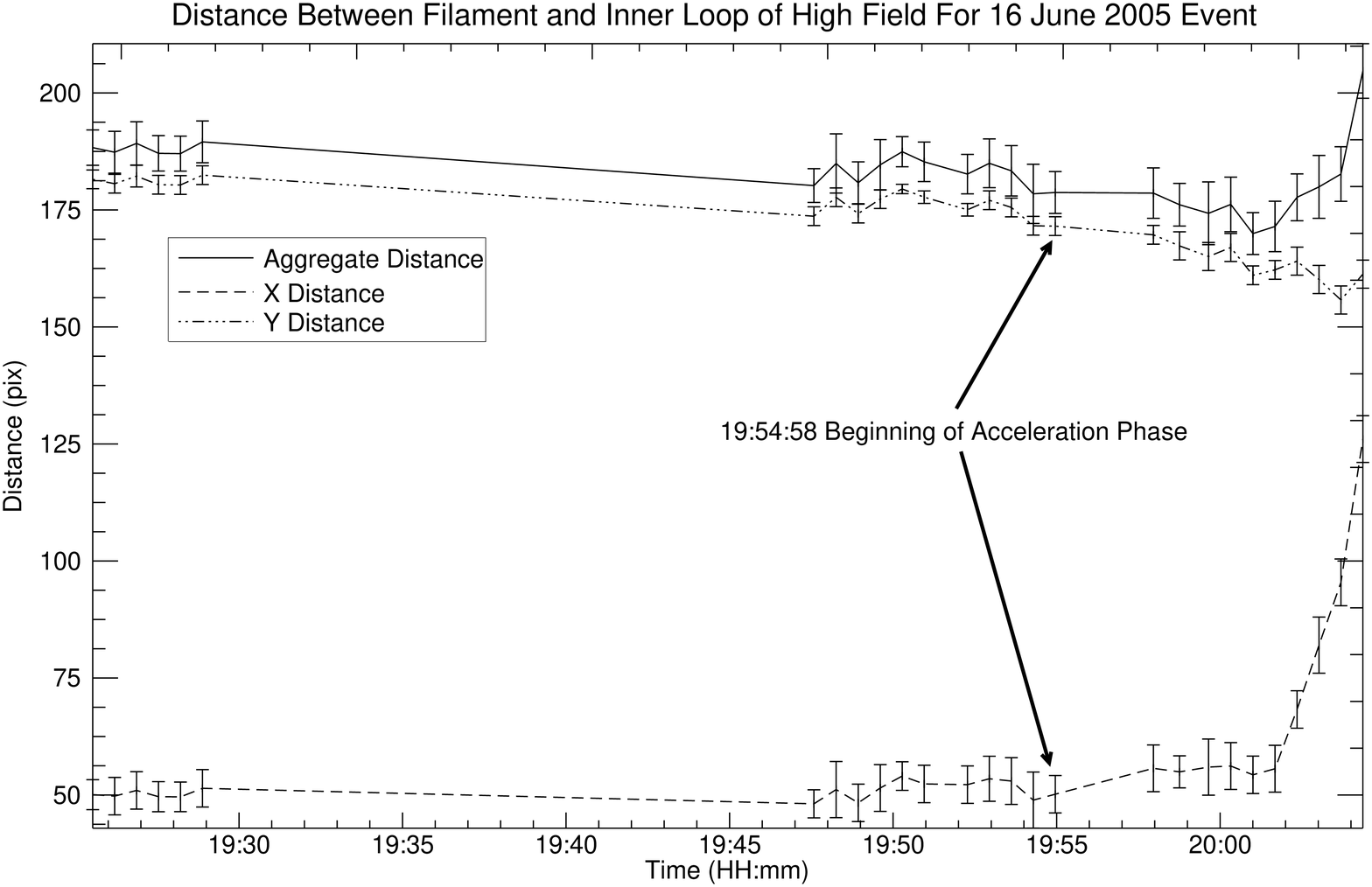}
\caption{\em }\label{fig:1c}
\end{figure}

\begin{figure}[ht!]
\epsscale{.65}
\plotone{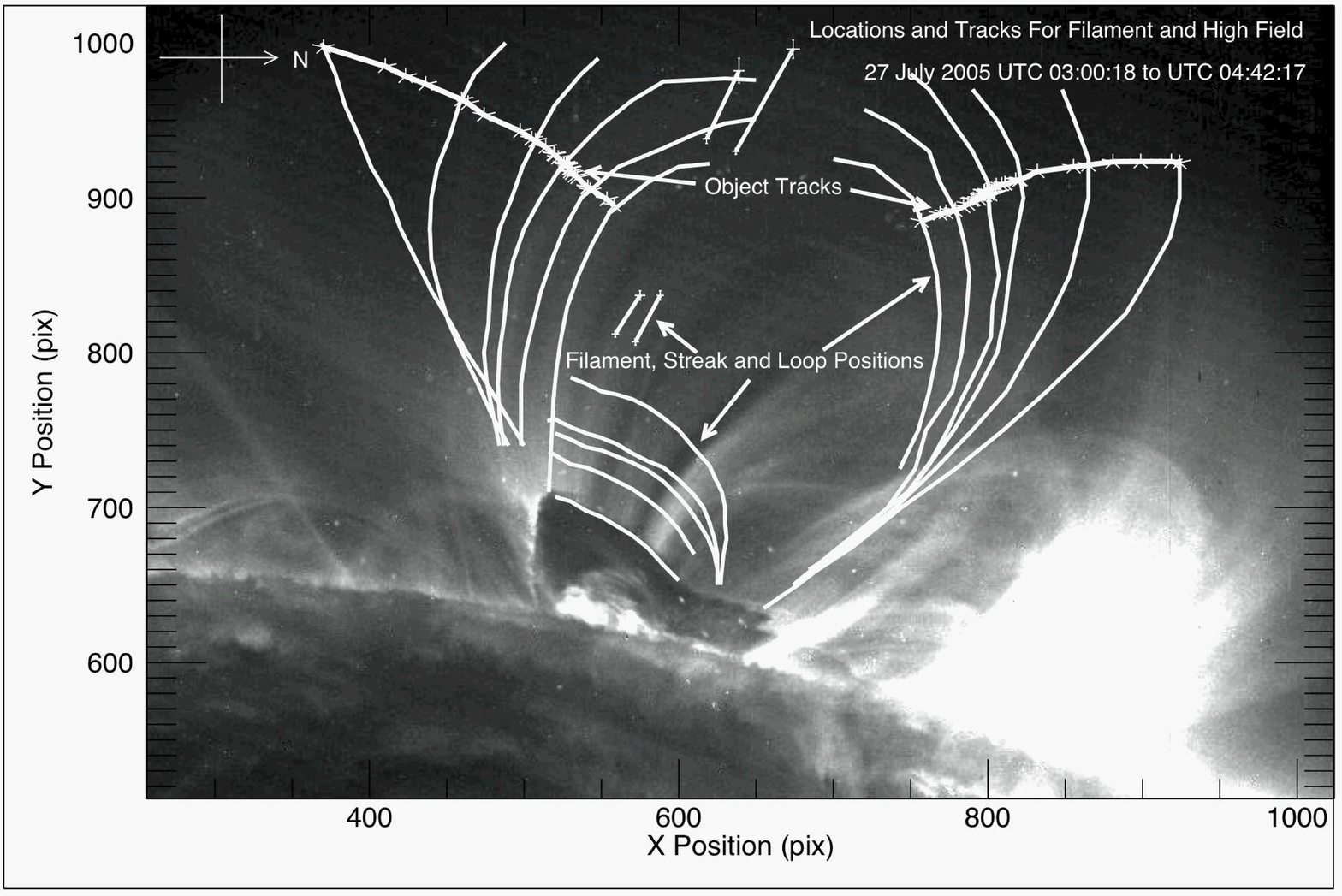}
\caption{\em }\label{fig:2a}
\end{figure}

\begin{figure}[ht!]
\epsscale{.65}
\plotone{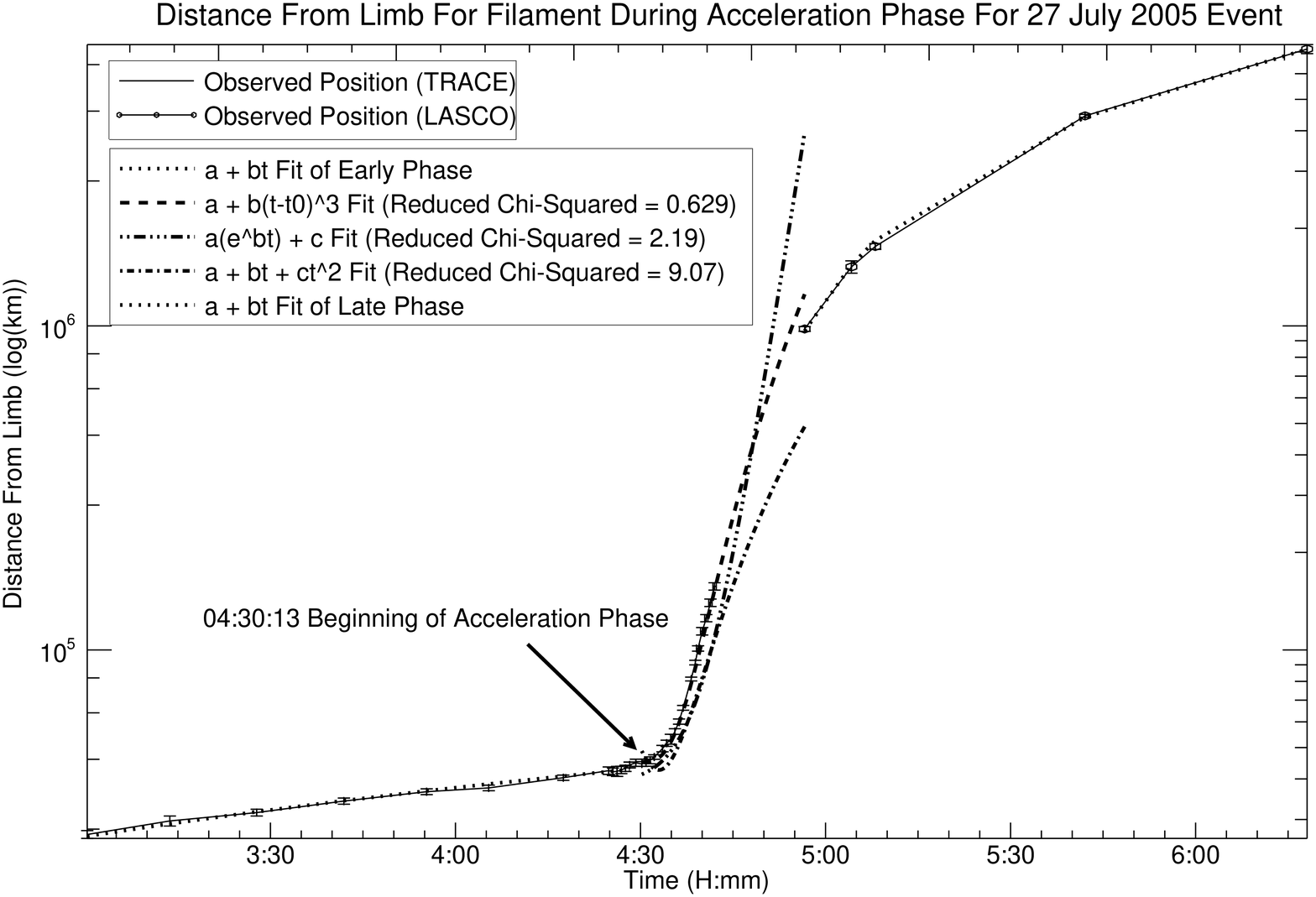}
\plotone{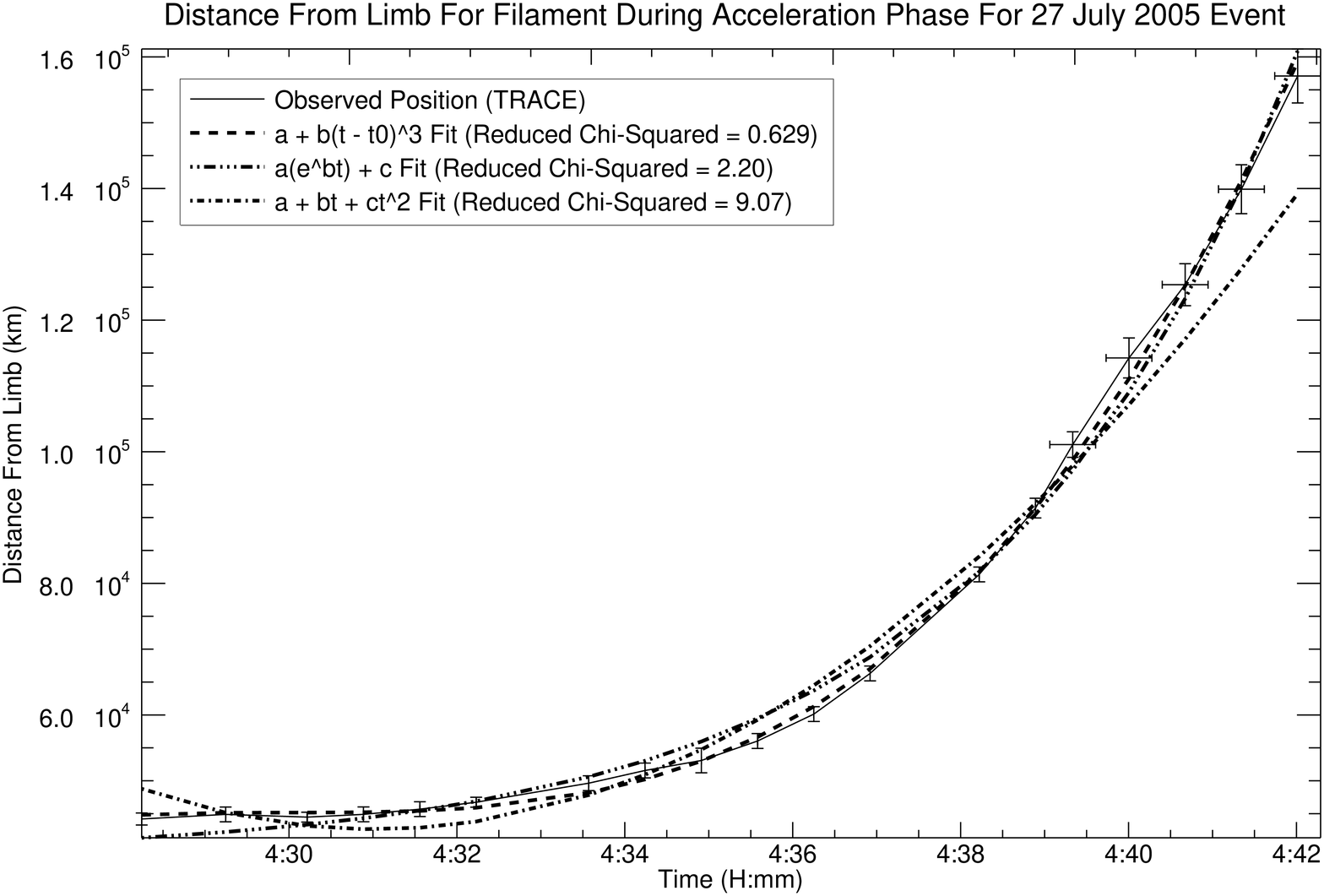}
\caption{\em }\label{fig:2b}
\end{figure}

\begin{figure}[ht!]
\epsscale{.65}
\plotone{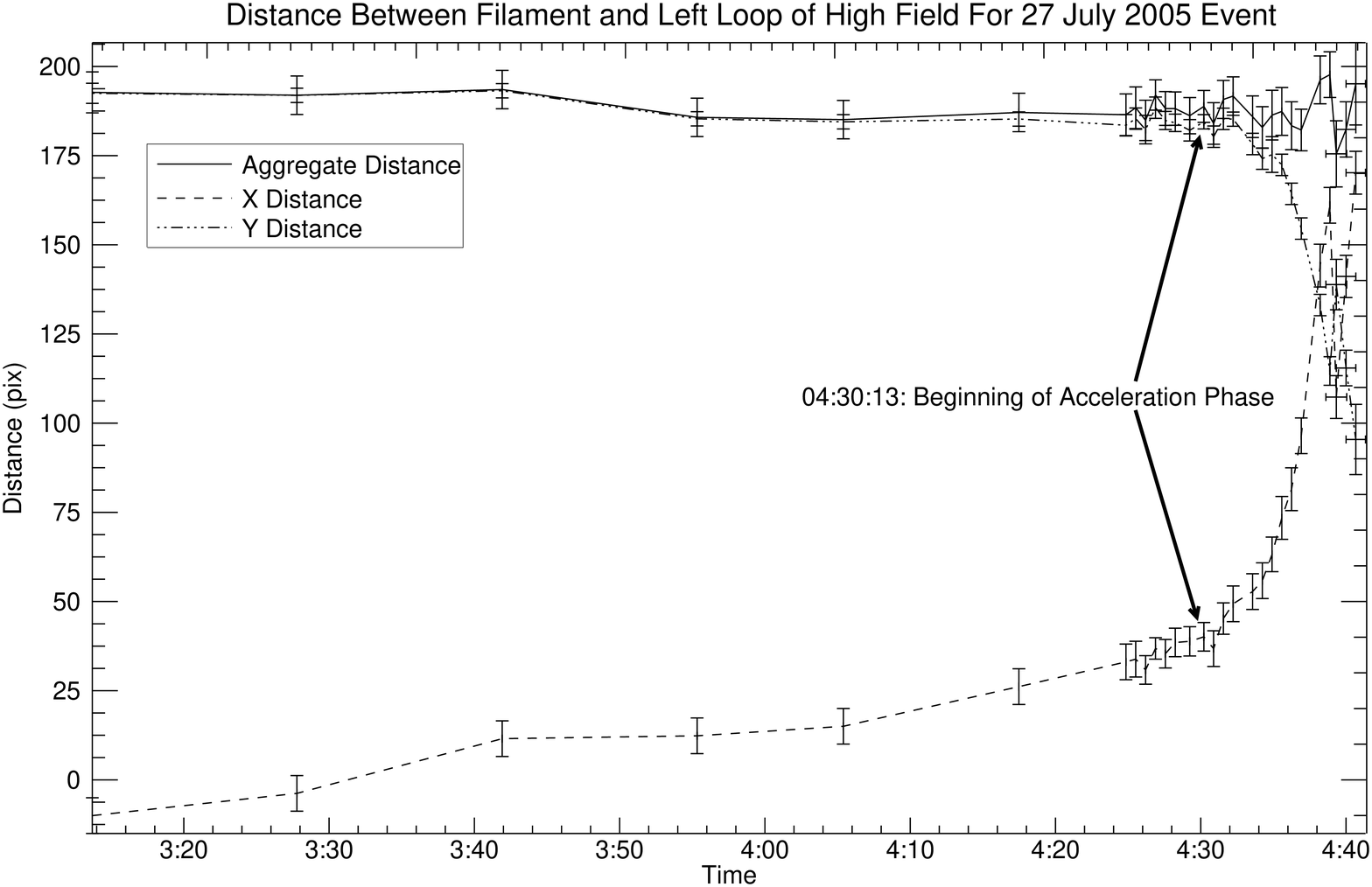}
\plotone{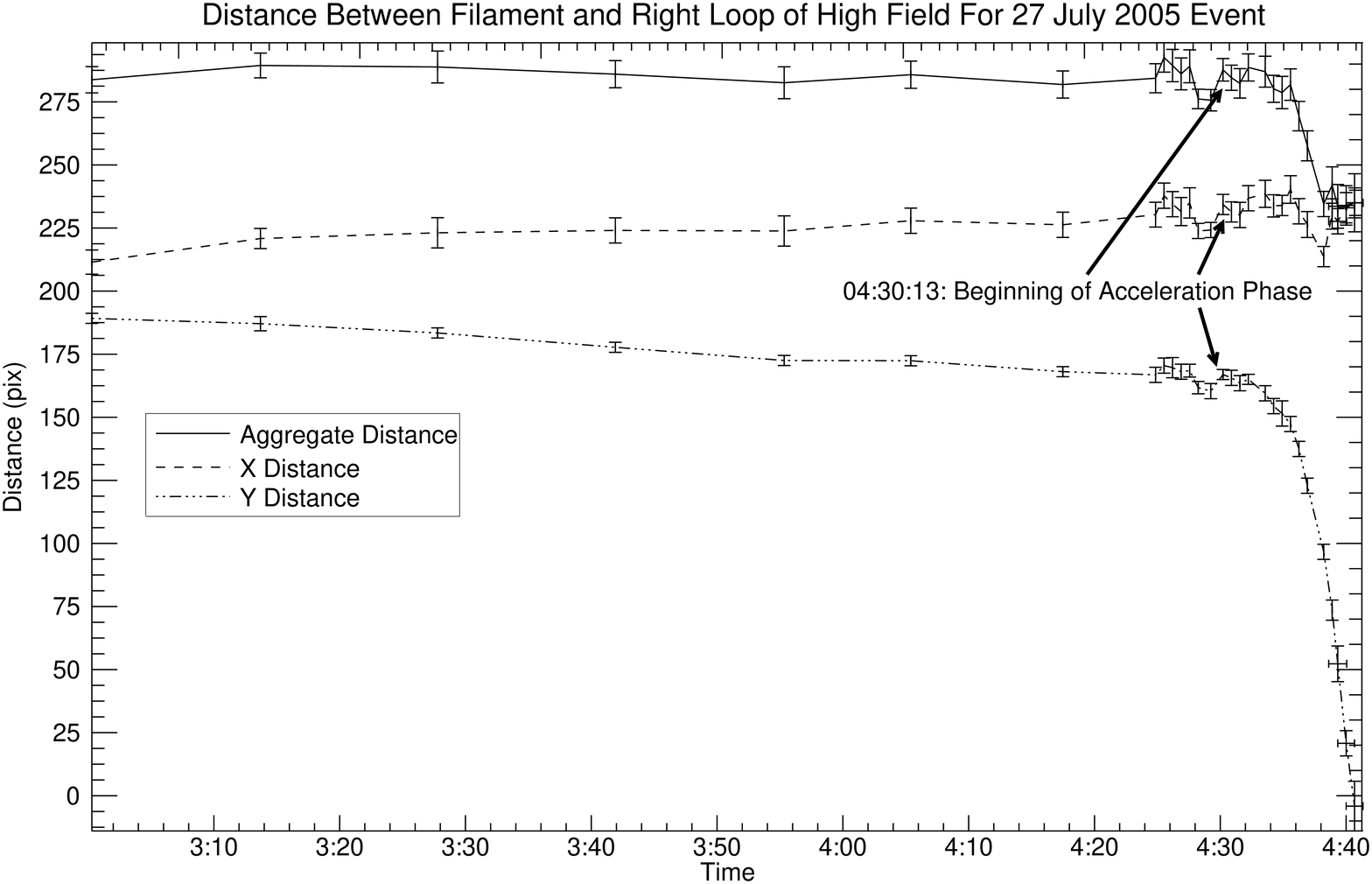}
\caption{\em }\label{fig:2c}
\end{figure}

\begin{figure} 
\epsscale{.65}
\plotone{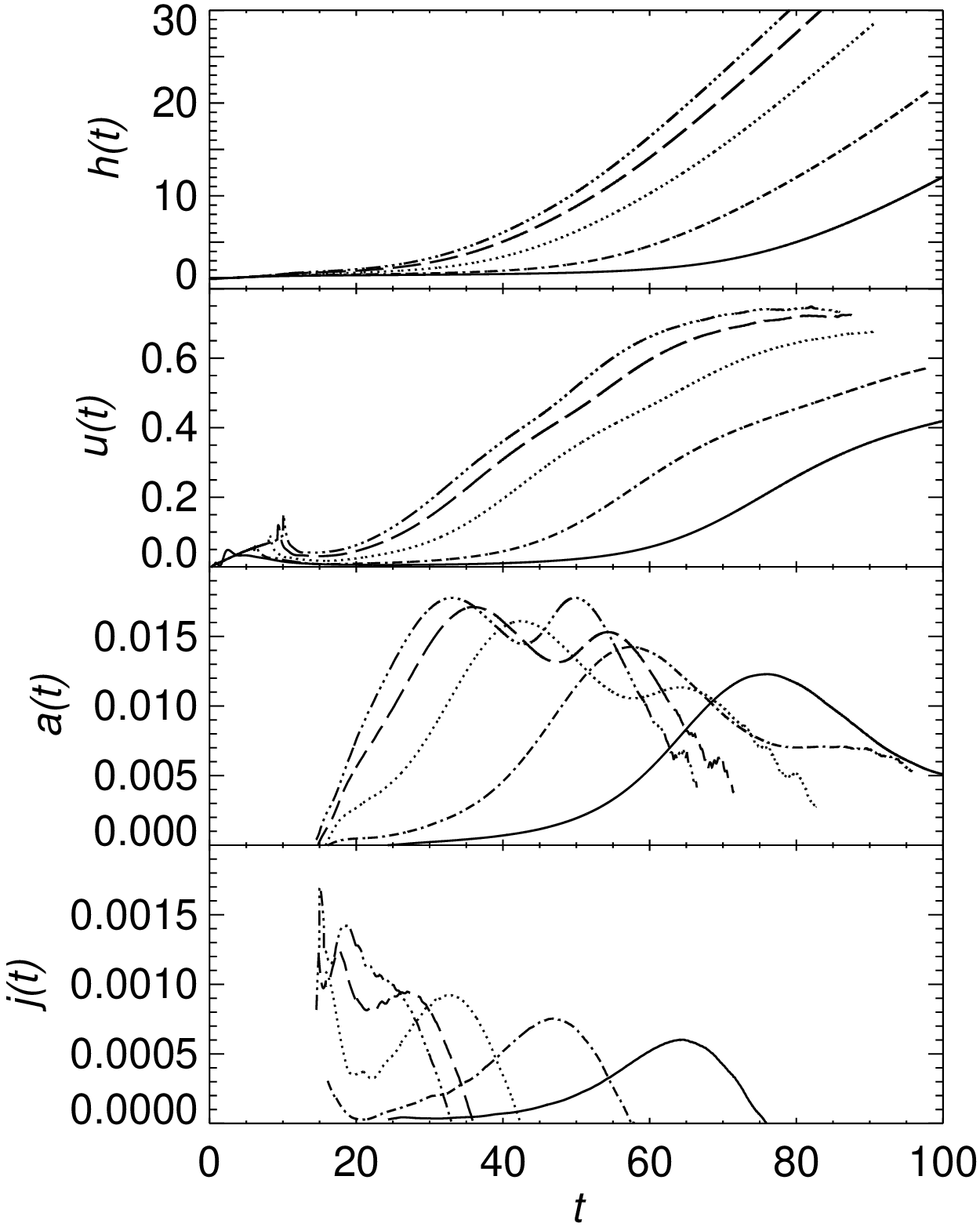}
\caption[]
{}
\label{fig:TIseries}
\end{figure}

\begin{figure} 
\epsscale{.65}
\plotone{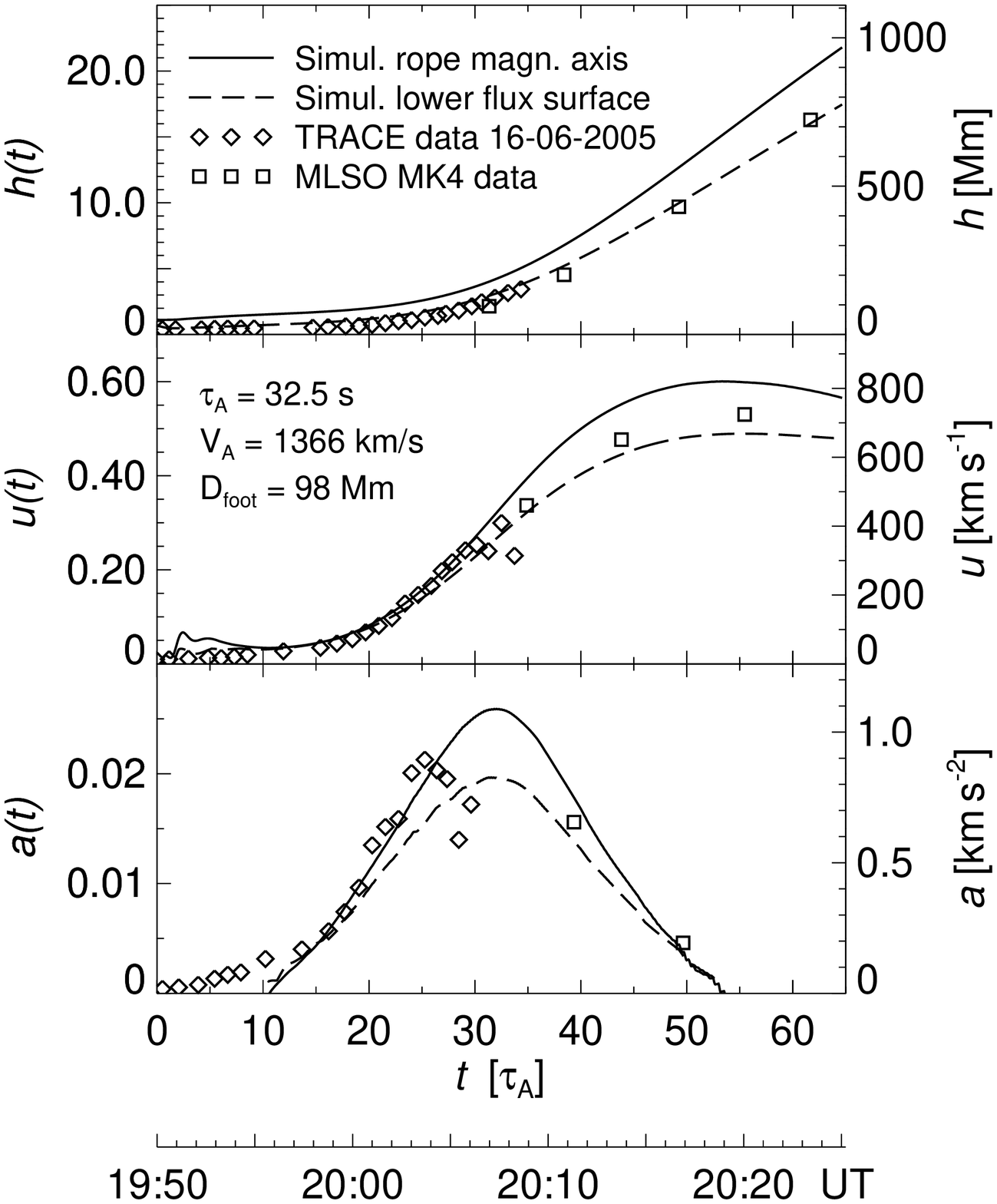}
\caption[]
{}
\label{fig:constant_jerk_TI}
\end{figure}

\begin{figure}[ht!]
\epsscale{.5}
\plotone{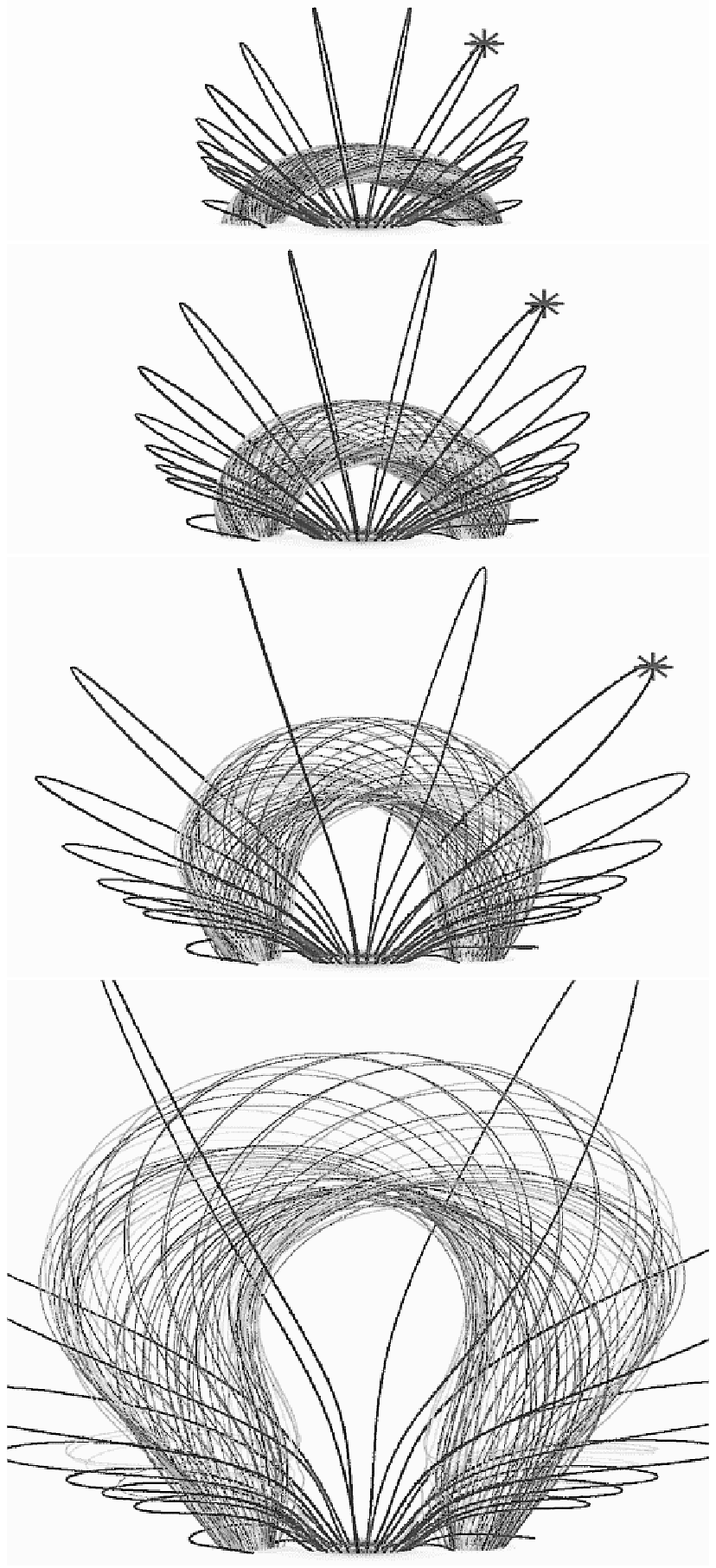}
\caption{\em }\label{fig:simseq}\label{fig:TI_4frame_rendering}
\end{figure}

\begin{figure} 
 \centering
 \includegraphics[width=3.25in]{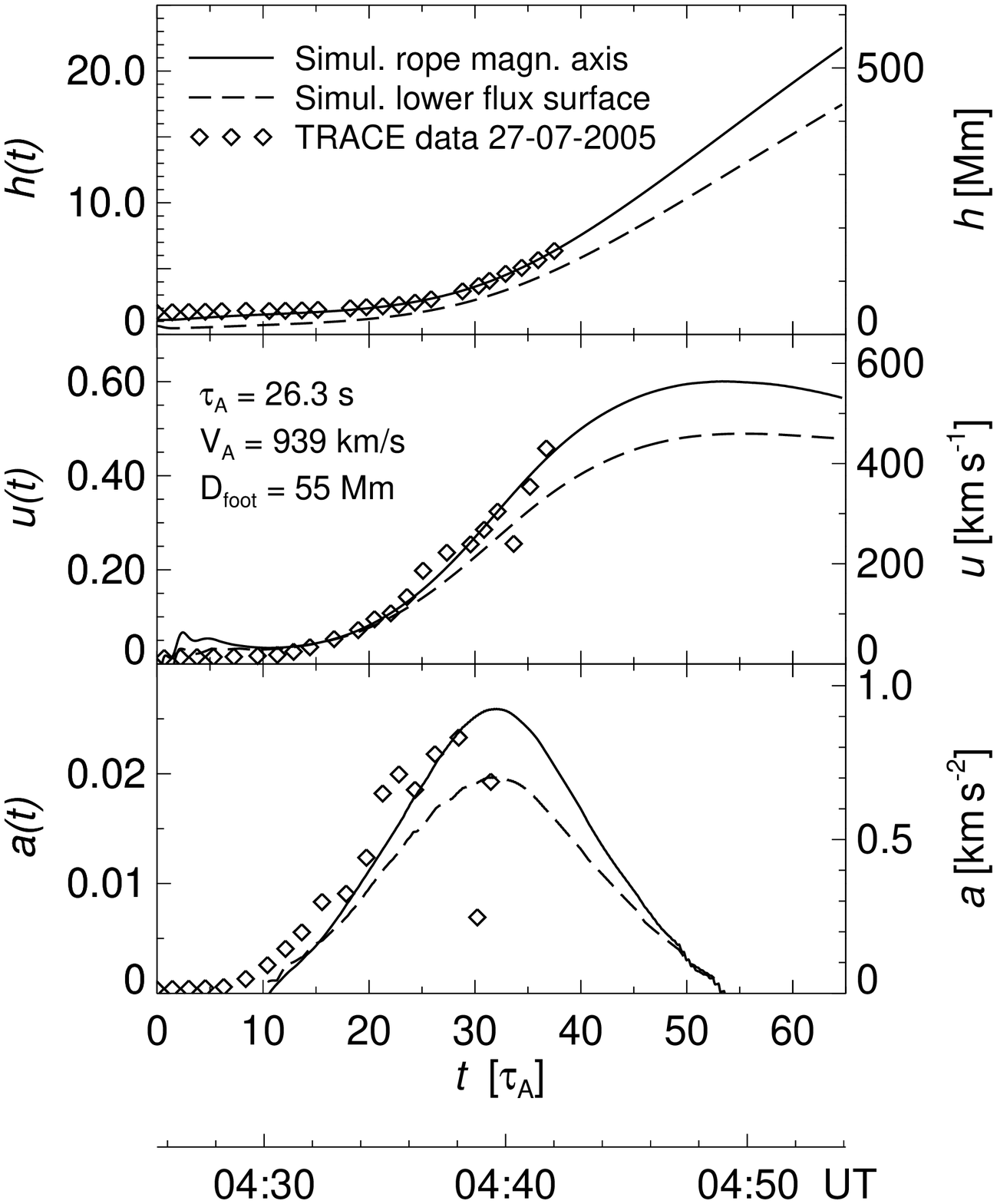}
\caption[]
{}
\label{fig:scaling20050727}
\end{figure}

\begin{figure} 
 \centering
 \includegraphics[width=2.5in]{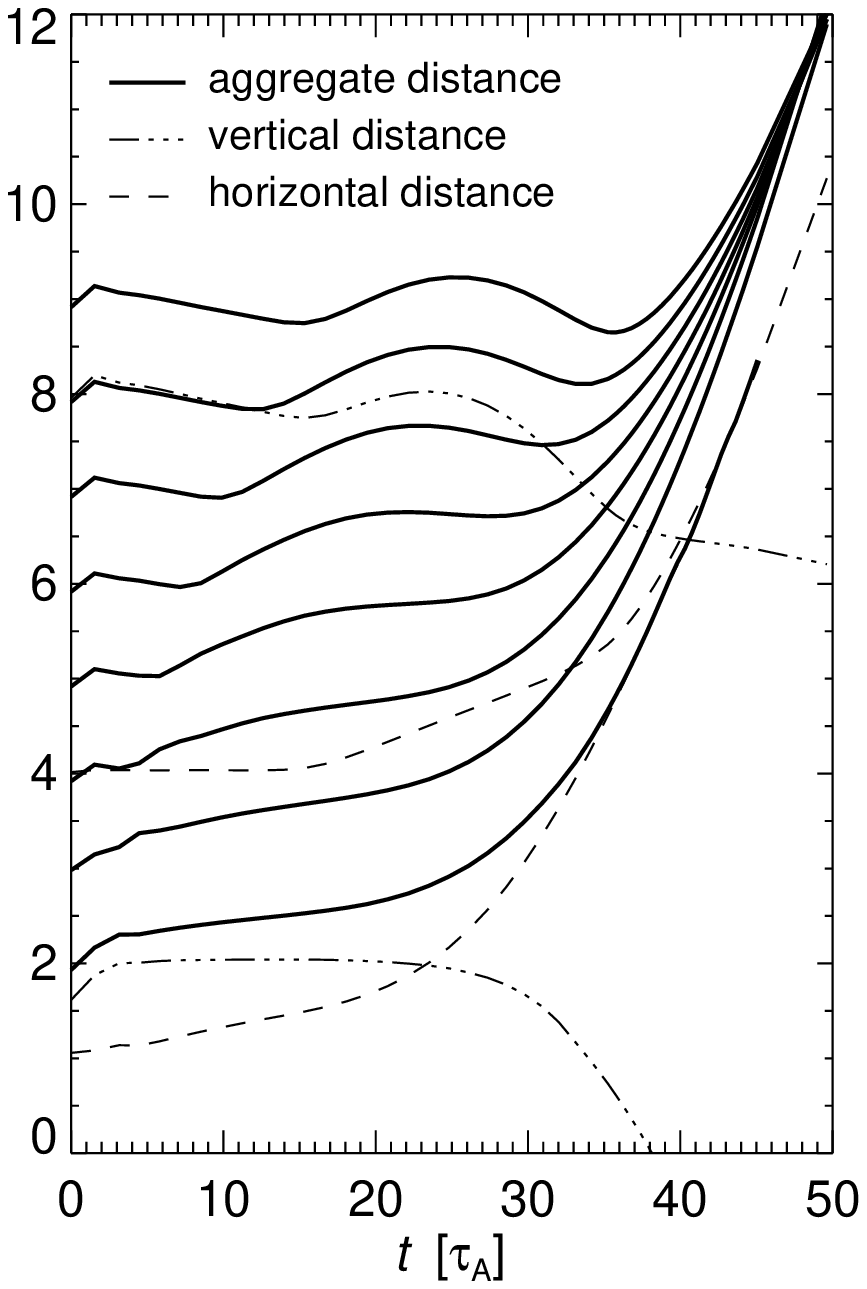}  
\caption[]
{}
\label{fig:OverlyingLoops}
\end{figure}

\end{document}